\documentclass[runningheads]{llncs}
\usepackage{graphicx}
\usepackage{xcolor}
\usepackage{multirow}
\usepackage{multicol}
\usepackage{mathtools}
\usepackage{amsfonts}
\usepackage{wrapfig}
\usepackage{array}
\usepackage{makecell}
\usepackage[math]{cellspace}
\usepackage{tablefootnote}
\usepackage{booktabs}
\usepackage{changepage}
\usepackage{newverbs}
\usepackage{pgffor}
\usepackage{listings}
\usepackage{bold-extra}
\usepackage[linesnumbered,ruled]{algorithm2e}
\usepackage{longtable} 
\usepackage{setspace}
\usepackage{hyphenat}
\usepackage{verbatim}
\usepackage{adjustbox}
\usepackage{tabularx}
\usepackage{floatrow}

\usepackage{tikz}
\usetikzlibrary{shapes.geometric}
\usetikzlibrary{shapes.arrows}
\usetikzlibrary{decorations.pathreplacing}
\usetikzlibrary{calc}
\usetikzlibrary{positioning}

\usepackage[colorlinks = true,
            linkcolor = ,
            anchorcolor = ,
            citecolor = ,
            filecolor = ,
            menucolor = ,
            runcolor = ,
            urlcolor  = blue,]{hyperref}

\newfloatcommand{capbtabbox}{table}[][1.0\FBwidth]

\newcommand{\lit}[1]{\texttt{#1}}
\newcommand{\itt}[1]{\textit{#1}}

\newcommand{\goal}[1]{\left\langle\begin{smallmatrix}#1\end{smallmatrix}\right\rangle}
\newcommand{\goalbag}[1]{\itt{\textbf{#1}}\,}

\def\mcolour#1#{\@mcolour{#1}}
\def\@mcolour#1#2#3{%
  \protect\leavevmode
  \begingroup
    \color#1{#2}#3%
  \endgroup
}

\newcommand{\algoName}{Child-Generator-Deque-Search}
\newcommand{\algoShort}{CGDS}

\newif\ifanonymous

\begin{document}
%
\title{Cain: Automatic Code Generation for Simultaneous Convolutional Kernels on Focal-plane Sensor-processors}
\titlerunning{Cain: Convolutional Filter Compiler for Focal-plane Sensor-processors}
%
%
\ifanonymous
\else
\author{Edward Stow\inst{1} \and
Riku  Murai\inst{1} \and
Sajad Saeedi\inst{2} \and Paul H J Kelly \inst{1}}
\authorrunning{E. Stow et al.}
%
\institute{Dept of Computing,
Imperial College London \\ 180 Queens Gate, London, SW7 2AZ, United Kingdom\\
\email{\{edward.stow16, riku.murai15, p.kelly\}@imperial.ac.uk}\\
\and
Dept of Mechanical and Industrial Engineering, Ryerson University, \\
350 Victoria Street, Toronto, Ontario, M5B 2K3, Canada\\
\email{s.saeedi@ryerson.ca}
}
\fi

\maketitle              
\setcounter{footnote}{0} 
\begin{abstract}
Focal-plane Sensor-processors (FPSPs) are a camera technology that enable low power, high frame rate computation, making them suitable for edge computation. Unfortunately, these devices' limited instruction sets and registers make developing complex algorithms difficult. In this work, we present Cain\footnote{Available at \url{https://github.com/ed741/cain}} -- a compiler that targets SCAMP-5, a general-purpose FPSP -- which generates code from multiple convolutional kernels. 
As an example, given the convolutional kernels for an MNIST digit recognition neural network, Cain produces code that is half as long, when compared to the other available compilers for SCAMP-5.

\end{abstract}

\section{Introduction}

Real-time computer vision applications are currently bound to traditional camera sensors that transfer each pixel at each frame to a host where it is processed. This requires high-performance buses between the sensors and hosts, especially where high frame-rates are required. A self-driving car may need to receive new information for every 1cm travelled to be vigilant of unexpected scenarios, so at 80 km/hr a frame rate of 2222 Hz would be required. A 2 mega-pixel camera, with 10-bit pixel depth, running at such a frame rate, requires a bus capable of 45.6 Gbit/s --- which is currently only possible with devices such as a PCI-e x8 Gen3 interface~\cite{pcieCam}. 
For many applications, however, streaming data at such volumes is too demanding -- both in power and computation time -- hence requiring an alternative solution.

Codesign of hardware and software for computer vision applications is an emerging research field to address the limitations of conventional systems \cite{8436423}. Focal-plane Sensor-processors (FPSPs) are a promising avenue for reducing the data transfer between the camera and the processing unit. FPSPs, often synonymous with Cellular Processor Arrays (CPAs) and Pixel Processor Arrays (PPAs), perform processing on the sensor chip itself and are often designed for tasks which require high frame rates or low latency~\cite{fpsp1}. The principle behind them is that a small processor is embedded directly with each pixel of the sensor. While FPSPs come in various forms for specific applications, we in this paper we explore a general-purpose fine-grain architecture SCAMP-5~\cite{scamp1}, but one can imagine alternatives that could be designed for various use cases.

One of the most widely used methods for image analysis is convolution kernels.
From edge detection using Sobel filters to document recognition using Convolutional Neural Networks~\cite{lecun1998gradient}, convolutional kernels are the foundation for many complex computer vision applications.  Traditionally, application of the convolutional kernels to the image data occurs on a CPU, but more recently GPUs and FPGAs are used to accelerate the computations in parallel~\cite{abadi2016tensorflow},~\cite{chen2016eyeriss}. Several systems have been designed to optimise the processing of convolutional kernels on GPUs and FPGAs, leading to a vast array of techniques to reduce the number of operational cycles needed to apply kernels to input data. 
While this significantly increased throughput, these methods are still bounded in latency as the image must make its way from the camera through to the host system.
As for FPSPs, the ability to process the data on the focal plane enables the kernels to be applied to the image data at very low latency. Furthermore, the unique ability to select the data which is transferred from the device to the host reduces the data volume, which allows for high frame rates. However, the technology is comparatively new. By design, they offer novel ways to interact with the data, and while work has been done to provide a Domain-Specific-Language and associated tools to program such hardware~\cite{martel}, there has been less work done so far to produce code generation systems to make efficient use of their architectural features when applying convolutional kernels in particular. 

One such system that does exist, however, is AUKE \cite{TomD}. 
Given an $N \times N$ convolutional kernel, AUKE's reverse-split algorithm generates code for SCAMP-5 which applies the kernel efficiently to the captured image on the focal-plane using analogue computation. AUKE is, however, limited to compiling just a single convolutional kernel at a time using a reduced instruction set that omits the more powerful instructions available in SCAMP-5.

In this work, we present an improved alternative to AUKE, with the ability to produce code for applying multiple convolutional kernels at a time. 
The problem is presented as a dynamic graph search problem in which we must efficiently generate and traverse possible processor states to find a path that describes the relevant convolutional computation. By incorporating instruction selection and instruction scheduling into the core of search process, we enable the use of more novel features of CPA architectures than AUKE is able to use.
By optimising the code for multiple kernels simultaneously, common sub-expressions between kernels can be exploited and produced only once rather than for each kernel. This reduces the computational expense of applying the kernels, enabling applications to run at a faster frame rate. 

The primary objective of this work is to push the boundary of code generation for FPSP devices through simultaneous kernel optimisation.
We offer the following contributions:
\vspace{-0.2em}
\begin{itemize}
    \item Cain: A code generation algorithm which effectively makes use of common sub-expressions across filters consisting of multiple convolutional kernels. Our graph search strategy -- which enables Cain to efficiently search large graphs -- combines instruction scheduling, instruction selection and register-allocation constraints into the core of the search to make better use of specific hardware capabilities in SIMD processors.
    \item We show how this search can be tractable for problems of interest through a problem formulation based on AUKE's multi-set--of--Atoms problem representation, combined with a ranking heuristic and a hybrid graph-generator--graph-search exploration strategy.
    \item We show how this approach allows flexible exploitation of hardware capabilities (such as three-operand adds and multi-step shifts), and generates very efficient use of additions to avoid multiplies.
    \item Evaluation of the effectiveness of Cain on the SCAMP-5 Focal-plane Sensor-processor. We compare against AUKE and test the effectiveness of simultaneous kernel optimisation. We conclude by exploring how our simultaneous kernel optimisation extends to future devices with more registers per pixel.
\end{itemize}\vspace{-0.2em}

The remainder of the paper is organised as follows.
Section~\ref{sec:background} describes the SCAMP-5 and its instruction sets,
Section~\ref{sec:cain} explains our proposed code generation algorithm Cain, and in Section~\ref{sec:evaluation} detailed comparison is made between Cain and AUKE, together with an evaluation of the effectiveness of simultaneous kernel optimisation. Section~\ref{sec:related_work} reviews the related work AUKE in detail.
Finally, Section~\ref{sec:conclusion} concludes our work, with a discussion about potential future research.

\begin{figure}[t]
    \centering
     \scalebox{0.5}{
  \begin{tikzpicture} [
    auto,
    node distance = 1cm and 2cm,
    decision/.style = { diamond, draw=blue, thick, fill=blue!20,
                        text width=5em, text badly centered,
                        inner sep=1pt, rounded corners },
    block/.style    = { rectangle, draw=blue, thick, 
                        fill=blue!20, text width=10em, text centered,
                        rounded corners, minimum height=2em },
    line/.style     = { draw, thick, ->, shorten >=2pt },
    bline/.style     = { draw, thick, shorten >=2pt, blue },
  ]

    \node [text centered] (kernel) {\textbf{Input Kernels}};        
    \node [block, below= of kernel] (aprox) {Goal Approximation};          
    \node [block, below= 2cm of aprox] (tsys) {Configurable Traversal System}; 
    \node [decision, right= of tsys] (done) {Explore Node};       
    \node [block, right=2.2cm of done] (genpairs) {Generate Next Goal Pair}; 
    \node [block, right= of genpairs] (elem) {Apply Instruction\\ In Reverse};   
    \node [block, above=of done] (reg) {Register Allocation};      
    \node [block, right=of reg] (code) {Code Generation}; 
    \coordinate [below= 2cm of tsys] (null1) {};
    \coordinate [below= 2cm of elem] (null2) {};
    \coordinate let \p1 = (done), \p2 = (null1) in coordinate (null3) at (\x1,\y2) {};

  \begin{scope} [every path/.style=line]
    \path (kernel) -- node [left] {Matrices of Coefficients}  (aprox);
    \path (aprox)  --    node [left, align=center] {Final-Goals (root node)} (tsys);
    \path (tsys)    --  node[align=center] {Parent\\ Node}   (done);
    \path (done)   -- node [left,align=center] {Specific\\ Instructions} node [near start, right] {\textbf{Initial-Goal Found}} (reg);
    \path (done)    --    node [midway] {\textbf{Otherwise}} (genpairs);
    \path (done) -- node [near start] {\textbf{Node Culled}} (null3) -- (null1) -- (tsys);
    \path (genpairs)   -- (elem);
    \path (elem) -- (null2) -- node [above,near start,align=center] {Parent Node and Child Node} (null1) -- (tsys);
    \path (reg) --    (code);
  \end{scope}
  
\end{tikzpicture}
}
    \caption{Cain System Overview. }
    \label{fig:CainSystemOverview}

\end{figure}
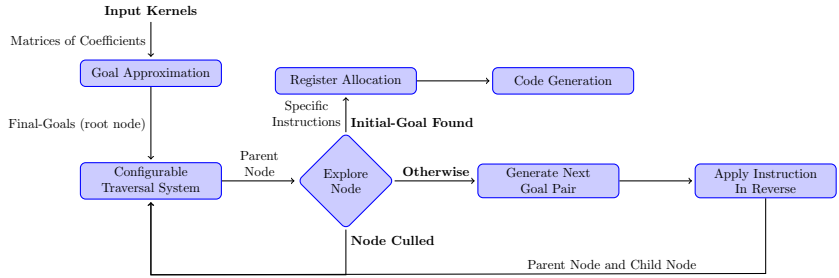

\begin{figure}[t]
    \centering
{
\newcommand{\gb}[1]{$\left\{ #1 \right\}$}
\newcommand{\cgoal}[3]{\goal{#1\\ #2\\ #3}}
\newcommand*\circled[1]{\tikz[baseline=(char.base)]{
            \node[shape=circle,draw=green,inner sep=1pt, font=] (char) {#1};}}
\scalebox{0.75}{
\begin{tikzpicture}[level distance=6em, font=\scriptsize,
    every node/.style = {inner sep=2pt, shape=rectangle, rounded corners, draw, align=center},
    level/.style={sibling distance = 3em + 7em/(#1),
  level distance = 1.1cm},
    fail/.style = {draw=red, fill=red!20},
    dots/.style = {inner sep=10pt, draw=none},
    missing/.style = {draw=none},
    line/.style     = { draw=green, thick, ->, shorten >=2pt},]
    \node (root) at (-3,0) {\gb{\cgoal{1}{2}{1}}}
        child { node {\gb{\cgoal{1}{2}{0},\cgoal{0}{0}{1}}}
            child { node {\gb{\cgoal{1}{2}{0},\cgoal{0}{1}{0}}}
                child { node [fail] {\gb{\cgoal{1}{2}{0},\cgoal{0}{0}{1}}} edge from parent node[left, draw=none]{\lit{mov()}} } 
                child { node [dots] {\huge\dots} }  edge from parent node[left, draw=none]{\lit{mov()}}
                    }
            child { node [dots] {\huge\dots} }  edge from parent node[above=0.1cm, draw=none]{\lit{add()}}
                }
        child { node {\gb{\cgoal{0}{1}{1},\cgoal{1}{1}{0}}}
            child { node [missing] {} edge from parent[draw=none] }    
            child { node {\gb{\cgoal{0}{1}{1}}}
                child { node {\gb{\cgoal{0}{1}{0},\cgoal{0}{0}{1}}}
                    child { node {\gb{\cgoal{0}{1}{0}}}  edge from parent node[left, draw=none]{\lit{mov()}}
                            }
                    child { node [dots] {\huge\dots}  }  edge from parent node[left, draw=none]{\lit{add()}}      
                        }    
                child { node [dots] {\huge\dots} }  edge from parent node[left, draw=none]{\lit{mov()}}      
                    }
            child { node [dots] {\huge\dots} } edge from parent node[left, draw=none]{\lit{add()}}       
                }
        child { node [dots] {\huge\dots}
                }
    ; 
    
    \node[draw=none] at (3,-3) {%
            \large
            \begin{varwidth}{5cm}
            $
            \begin{array}{p{1em}c|p{3cm}}
                   & \text{Step:} & \text{Instruction:}\\
                   \hline
                 1 & \circled{7} & \lit{mov(B,A,south)} \\
                 2 & \circled{6} & \lit{add(A,A,B)} \\
                 3 & \circled{5} & \lit{mov(B,A,north)} \\
                 4 & \circled{4} & \lit{add(A,A,B)}
            \end{array}
            $
        \end{varwidth}
            };
    \begin{scope} [every path/.style=line]
        \path (root) edge[bend right] node[above] {1} (root-1);
        \path (root-1) edge[bend right=70] node[above] {2} (root-1-1);
        \path (root-1-1) edge[bend right=70] node[above] {3} (root-1-1-1);
        \path (root) edge[bend left=90] node[left] {4} (root-2);
        \path (root-2) edge[bend left=90] node[left] {5} (root-2-2);
        \path (root-2-2) edge[bend left=70] node[left] {6} (root-2-2-1);
        \path (root-2-2-1) edge[bend left=70] node[left] {7} (root-2-2-1-1);
    \end{scope}
    \begin{scope}[every path/.style= {draw=red, ->, double distance=2pt }]
        \path (root-1-1-1) edge [bend right=65] (root-1);
    \end{scope}
    
\end{tikzpicture}
}
}
    \caption{Graph showing how Cain might search a simplified 1-dimensional problem using \algoShort. Numbered steps show the order that the paths are explored with child nodes generated the first time a search step starts at a parent node. Nodes are checked for being the Initial-Goal when pointed too. The red node, and edge, correspond to a dead-end where a duplicate node has been found at a higher cost than previously seen and so the node is not traversed further. We see a path to the Initial-Goal is found after 7 steps, and the code produced by this path is presented on the right. The \lit{mov()} instruction in step 5 exploits a common sub-expression such that the two Goals in its output Goal-Bag are produced together, thus shortening the code.}
    \label{fig:CainSearchGraph}
    \vspace{-0.5em}
\end{figure}

\section{Background: SCAMP-5 Focal-plane Sensor-processor}\label{sec:background}
In this section, we discuss the capabilities of the next generation camera technology SCAMP-5, and give an overview of the functionality used by Cain.

SCAMP-5 has been demonstrated in many different computer vision applications, ranging from Visual Odometry systems~\cite{murai2020bit},~\cite{bose_visual_2017},~\cite{debrunner2019Multiprog}, an end-to-end neural sensor which performs learnt pixel exposures~\cite{DBLP:journals/pami/MartelMCDW20}, to Convolutional Neural Networks~\cite{wong2020analognet},~\cite{bose2019camera}. Its distinctive ability to perform computation on the focal-plane reduces power consumption and data transfers, making the device promising for edge computation.

The SCAMP-5 architecture is a general-purpose fine-grain SIMD FPSP \cite{scamp2}. It has a $256\times 256$ pixel array, and along with each pixel is a small Processing Element (PE). All 65,536 processors execute the same instruction at one time. In addition to 14 binary registers, each PE has analogue registers \itt{A} through to \itt{F} as well as a \itt{NEWS} register. Each PE can also address an \itt{XN}, \itt{XE}, \itt{XS}, and \itt{XW} register that is actually that PE's respective neighbours' \itt{NEWS} registers. Each PE uses an analogue bus to link its available analogue registers, and because values are stored as charge; analogue arithmetic is done directly on the bus that connects the registers rather than on a separate arithmetic unit.

Instructions in the architecture control how register values are let into and out of the bus with the caveat that values are inverted due to the nature of the analogue electronics. Each macro instruction like \lit{add}, \lit{sub}, and \lit{mov} are made of multiple bus instructions that create the desired behaviour, where the $\lit{bus}n (w_1,..w_n, r_0..r_k)$ instruction has the general rule that the values of registers $r_0 .. r_k$ are summed up, negated, and divided equally between the $n$ receiving-registers $w_1..w_n$. Since a bus operation directly controls which registers are opened to the PE's common analogue bus, a register may only appear once in each \lit{bus} instruction. Each bus instruction also incurs significant noise and error factors, especially for \lit{bus2} and \lit{bus3}~\cite{scamp3}.

Macro instruction arguments are written as if they are assignment statements.
For example; the macro instruction \lit{add(A, B, C)} means $A := B+C$ and is made up of two bus instructions: \lit{bus(NEWS, B, C)} meaning the \itt{NEWS} register now contains the value of $-(\itt{B}+\itt{C})$; and then \lit{bus(A, NEWS)} so that register \itt{A} contains $\itt{B}+\itt{C}$. 
We can see here that the \lit{add} instruction has additional constraints, such that the two operands cannot be the same register, and that the \itt{NEWS} register is overwritten, and left containing $-(\itt{B}+\itt{C})$ as a side effect. 
When using macro instructions, we restrict the registers to \itt{A} to \itt{F}, and allow the macros themselves to make use of the \itt{NEWS} and neighbouring \itt{NEWS} registers for us by means of a direction value. We use subscripts to denote the registers of neighbouring PEs.
For example: \lit{mov2x(A, B, north, east)} computes $A:=B_{\text{north},\text{east}}$ in two \lit{bus} instructions: \lit{bus(XS, B); bus(A,XE)}. The first means that 
$\itt{XS}_{\text{north},\text{east}} := B_{\text{north},\text{east}} $ which is equivalent to $ \itt{NEWS}_{\text{east}} := B_{\text{north},\text{east}}$ and then the second instruction means $A := \itt{XE} = \itt{NEWS}_{\text{east}} \implies A = B_{\text{north},\text{east}}$.

While interesting uses of the \lit{bus} instructions exist, allowing adding and subtracting from neighbouring PEs, individual macro instructions are still highly restricted in comparison to most modern instruction sets. Only primitive analogue operations are available to each PE such as: Move, Add, Subtract, Divide by two, and to acquire the value from the sensor \cite{scamp3}. The lack of a multiplication instruction means the problem of generating convolutional filter code for SCAMP-5 builds on the theory of multiplier-free FIR filters~\cite{CHANDRA2016212}.

The chip has been shown to be capable of operating at 100,000 FPS, largely because it is not limited by the speed of an output bus to transfer all the pixel data~\cite{scamp1}. Instead of only offering an analogue or digitally encoded output of all pixels at a time, like traditional camera sensors, the SCAMP-5 architecture allows binary outputs per pixel, and even event driven outputs. 
This allows each PE to come to a judgement on its input pixel data and fire its own event that sends the coordinates of the PE to the host; allowing information transfer without divulging the actual image.

The architecture uses an off-chip controller to manage the fetch-decode-execute cycle, with every pixel's processor receiving the same instruction, making it a single-instruction-multiple-data (SIMD) design. This has benefits in terms of simplicity and efficiency as none of the Processing Elements need to be able to fetch instructions for themselves. There is also provision for masking pixels such that only selected PEs execute instructions.

One important consideration to be made when using and designing algorithms related to the SCAMP-5 chip is noise introduced by the nature of the analogue computation. Every use of the 7 analogue registers introduces noise to the values stored. This makes finding optimal code to perform the convolutions ever more vital for accurate results. 

\section{Cain}\label{sec:cain}
Cain is a framework for compiling convolutional filters, designed to search through a configurable Cellular Processor Array (CPA) instruction set to find efficient code. A fundamental concept Cain uses is to only consider a single arbitrary PE in the CPA, and perform everything relative to it. This works for SIMD architecture like SCAMP-5 because every PE will be executing the same steps synchronously in parallel. The assumption we make when producing code is that the neighbours of our arbitrary PE will exist and so will have done the same work but at a relative offset in the input image. 
The aim is to search through the graph of possible Processing Element states in such a way that common sub-expressions in the given kernels are exploited and used to reduce the cost of any path from initial to final PE states. To do this Cain searches backwards, starting with a set of final kernels, these are the convolutional filter, and applying instructions in reverse to simplify the kernels until only the identity kernel\footnote{Single-entry matrix. Not to be confused with identity matrix} is left. Fig. \ref{fig:CainSystemOverview} shows a high level overview of this process. Searching backwards is a design choice that makes the search more effective because it means the aim at each step is to make what needs to be solved simpler than before. This means heuristics can be produced to always direct the search towards the identity kernel rather than a system of heuristics trying to accurately predict the path towards an arbitrary set of final Goals. We present this as a dynamic graph search problem because the size of the graph is intractable. Given the AnalogNet2 filter in Equation \ref{eq:AnalogNet2Filter}, Cain identifies 37163 potential children nodes in the first step alone. This can be reduced to 239 if we are willing to accept a less than exhaustive search of the solution space. This restriction is applied when the computational cost of computing the full set of children nodes is too high. 

\subsection{Definitions}
\label{sec:cain:goalDef}

This section provides an overview of notation and definition used in this paper. Cain is designed such that different definitions could be used without changing the fundamental search algorithm but the definitions we use here for SCAMP-5 are based largely on AUKE's, which provides an elegant way to conceptualise the convolutional kernels without multiplication.

\begin{example}{}
We will look at a simple example of how  a convolutional kernel is represented in Cain. Here we use AnalogNet2~\cite{wong2020analognet}\cite{analognet2} which is a CNN designed for SCAMP-5.
\vspace{-0.7em}
\begin{equation}
    \itt{AnalogNet2} =\begin{Bsmallmatrix}
    \frac{1}{4}
    \begin{bsmallmatrix}
    0 & 0 & 0\\
    -3 & 1 & 0\\
    -3 & 0 & 2
    \end{bsmallmatrix},&
    \frac{1}{4}
    \begin{bsmallmatrix}
    -4 & -1 & -1\\
    -1 & 2 & 0\\
    1 & 1 & 0
    \end{bsmallmatrix},&
    \frac{1}{4}
    \begin{bsmallmatrix}
    -1 & 2 & 0\\
    -1 & 1 & -3\\
    0 & -3 & 0
    \end{bsmallmatrix}
    \end{Bsmallmatrix}
    \label{eq:AnalogNet2Filter}
\end{equation}
\vspace{-2.1em}
\label{example:AnalogNet2}
\end{example}

Since SCAMP-5 does not have multiplication we must approximate the kernel and because it does have division-by-two instructions the natural approximation to make is to find the nearest integer multiple of $\frac{1}{2^d}$ for each coefficient in the kernel, given some number of divisions $d$. In our example we have already extracted the common denominator such that $d=2$ and this perfectly represents the kernel. The larger $d$ is, the larger the search space and complexity of the problem, so $d$ can be limited to allow an acceptable amount of approximation error such that the resulting program is shorter and computational expense of compiling it is reduced.

\begin{definition}{}
Let an Atom, denoted as $(x, y, z, \itt{sign})$, be a representation of $\frac{1}{2^d}$ of a pixel value at coordinate $x, y$, on the $z$th channel. $x, y$ are coordinates relative to the arbitrary PE and so also the centre of the kernel, and $z$ refers to an image input channel. The sign is used to negate the value if necessary.
\end{definition}
\begin{definition}{}
Let a Goal, denoted as $\{atom_1, atom_2,...\}$, be a multi-set of Atoms. The Goal represents an arbitrary kernel, however, scaled by $2^d$.  The aggregate of the values represented by each of the Atoms yields the same result as applying the scaled kernel. 
\end{definition}

Representing a convolutional kernel as a Goal is a convenient way to support multiply-free instruction set, such as SCAMP-5. One can simply view this as unrolling the multiply instruction into additions. Using Goals simply re-frames the problem by scaling everything by $2^d$, and approximating coefficients to the nearest number of Atoms. 
\begin{definition}{}
Let a Goal-Bag, denoted as $\{goal1, goal2,...\}$, be a multi-set of Goals. The Goal-Bag is used to capture the state of our arbitrary PE. This includes defining the Final-Goals, the set of convolution kernels we wish to compute; and the Initial-Goals, the set of Goals which the computation will start from. 
\end{definition}
Using these definitions of Goals and Atoms we see that the first kernel from Example \ref{example:AnalogNet2} can be represented by $G$
\vspace{-0.5em}
\begin{equation*}
K = \frac{1}{4}
    \begin{bsmallmatrix}
     0 & 0 & 0\\
    \textcolor{red}{-3} & \textcolor{blue}{1} & 0\\
    \textcolor{green}{-3} & 0 & \textcolor{brown}{2}
    \end{bsmallmatrix}, \quad  
    G = \begin{Bsmallmatrix}
    \textcolor{red}{(-1,0,0,-)},   & \textcolor{red}{(-1,0,0,-)},   & \textcolor{red}{(-1,0,0,-)},\\
    \textcolor{blue}{(0,0,0,+)},   & \textcolor{green}{(-1,-1,0,-)}, & \textcolor{green}{(-1,-1,0,-)},\\
    \textcolor{green}{(-1,-1,0,-)}, & \textcolor{brown}{(1,-1,0,+)}, & \textcolor{brown}{(1,-1,0,+)}
    \end{Bsmallmatrix}
\end{equation*}\vspace{-1em}

As our Goal notation is verbose, we provide a compact version that disambiguates Goals from kernels
\vspace{-0.5em}
\begin{equation}
    \begin{gathered}
    G = \goal{0&0&0\\-3&1&0\\-3&0&2} \implies \frac{1}{2^2}\begin{bsmallmatrix}
    0&0&0\\
    -3&1&0\\
    -3&0&2
    \end{bsmallmatrix} \star \text{Image Input}\label{eq:GoalConvolution}\\
    \text{\scriptsize where the $\star$ operator applies the left-hand convolutional kernel to the right-hand array}
    \end{gathered}
\end{equation}\vspace{-1em}

By repeating this for process the rest of the convolutional kernels in the AnalogNet2 filter, the Final-Goals Goal-Bag $\goalbag{FG}$ is produced:
\vspace{-0.5em}
\begin{equation}
\goalbag{FG} =
\begin{Bmatrix}
    \goal{
    0 & 0 & 0\\
    -3 & 1 & 0\\
    -3 & 0 & 2
    },&
    \goal{
    -4 & -1 & -1\\
    -1 & 2 & 0\\
    1 & 1 & 0
    },&
    \goal{
    -1 & 2 & 0\\
    -1 & 1 & -3\\
    0 & -3 & 0
    }
\end{Bmatrix}
\label{eq:kernelAproximation}
\end{equation}
Since, in our example, $d=2$; the Goal representation of the identity kernel ($G_\itt{ID}$) that makes up the Initial-Goals, is based on the approximation of the Final-Goals:
\vspace{-0.8em}
\begin{gather}
K_\itt{ID} = \frac{1}{4} \begin{bsmallmatrix} 0&0&0\\0&4&0\\0&0&0\end{bsmallmatrix} \implies G_\itt{ID} = \goal{
    0 & 0 & 0\\
    0 & 4 & 0\\
    0 & 0 & 0
    }\label{eq:exampleIDGoal}
\end{gather}
\vspace{-1em}

Moving a value around the processor array is expressed by translating every Atom of a Goal.
Addition and subtraction can be expressed by combining two Goals into one, making sure to cancel out positive and negative Atoms with the same coordinates. Since Cain searches backwards, we apply these operations in reverse. For 2-operand addition this means we take a Goal, $G$, that we wish to generate code for, then produce 2 new Goals that when added together produce $G$. 
Defining Goals as multi-sets of Atoms makes this process intuitive as we can simply split the Atoms between two Goals in every possible permutation (or fewer if we are willing to assume some are non-optimal, or willing to miss potentially better code for the sake of more efficient code generation). This definition also restricts the reverse search process since when splitting a Goal we cannot split an Atom.
To compute the red Atoms in $G$ naively, PEs must sum them and read this value from the west thus translating the Atoms eastward.

\subsection{Search Strategy}

Cain's reverse search algorithm works iteratively taking the state of an arbitrary PE, defined as a Goal-Bag:
\vspace{-0.5em}
\begin{equation}
    \goalbag{F} := \{G_1, G_2, G_2, G_3...\}\label{eq:defGoalBag}
\end{equation}\vspace{-1.5em}

This is a node in our search graph and represents the state we aim to achieve by executing the instructions that form a path from the initial-Goals to this node. In the search graph, nodes are generated dynamically as the graph is explored. Fig. \ref{fig:CainSearchGraph} shows a simplified view of how a graph might look as it is generated and searched.
We simplify the exploration such that in each iteration of the search algorithm we produce a Goal-Bag Pair of an $\goalbag{Uppers}$ Goal-Bag and a $\goalbag{Lowers}$ Goal-Bag as well as an instruction, with the following constraints:
\vspace{-0.5em}
\begin{align}
    (\goalbag{U},\goalbag{L}),inst = \itt{nextPair}(\goalbag{F})
    \text{ where } \goalbag{U} \subseteq \goalbag{F}, \; \goalbag{U} = inst(\goalbag{L})
\end{align}\vspace{-1.5em}

This is in contrast to AUKE's method, shown later in Equation \ref{eq:AukeReverseSplit}.
The new child node, $\goalbag{C}$, is then produced by applying the instruction in reverse using the following rule, with the instruction becoming an edge in the graph:
\vspace{-0.5em}
\begin{equation}
    \goalbag{C} = (\goalbag{F} \setminus \goalbag{U}) \cup \goalbag{L}
\end{equation}\vspace{-1.5em}

Following our AnalogNet2 example from Equation \ref{eq:kernelAproximation}, the first iteration of the search algorithm will start with $\goalbag{FG}$ and the Pair of Goal-Bags Cain produces is as follows:
\vspace{-0.5em}
\begin{gather}
    \goalbag{U} = \begin{Bsmallmatrix}
    \goal{-1&2&0\\-1&1&-3\\0&-3&0}
    \end{Bsmallmatrix},\quad
    \goalbag{L} = \begin{Bsmallmatrix}
    \goal{-1&2&0\\-1&1&0\\0&0&0},&
    \goal{0&0&0\\0&0&-3\\0&0&0},&
    \goal{0&0&0\\0&0&0\\0&-3&0}
    \end{Bsmallmatrix}\\
    \itt{inst} = \goalbag{U} \leftarrow \itt{add}(L_1, L_2, L_3)\\
    \goalbag{C} = \begin{Bmatrix}
    \goal{0 & 0 & 0\\-3 & 1 & 0\\-3 & 0 & 2},&
    \goal{-4 & -1 & -1\\-1 & 2 & 0\\1 & 1 & 0},&
    \goal{-1&2&0\\-1&1&0\\0&0&0},&
    \goal{0&0&0\\0&0&-3\\0&0&0},&
    \goal{0&0&0\\0&0&0\\0&-3&0}
\end{Bmatrix}\label{eq:exampleC}
\end{gather}\vspace{-1.0em}

The multi-set semantics here mean that if the Goals in $\goalbag{L}$ are all already part of $\goalbag{F}$ then the number of Goals to solve is reduced, and so by applying more pairs $(\goalbag{U},\goalbag{L})$ we traverse the graph of Goal-Bags, until we reach the initial-state, where the only Goal in the Goal-Bag is the identity Goal. In our example (Equation \ref{eq:exampleC}) we see that the sub-expression of 3 negative Atoms is reused in $C_4$ and $C_5$ since applying a $\lit{mov2x}$ next could eliminate $C_5$ from $\goalbag{C}$. There is also further potential to reuse this by how we split $C_1$. Once the initial Goal-Bag is found the path from the initial Goal-Bag back to the Final-Goals becomes the list of instructions that form our generated program. 

\begin{wrapfigure}[16]{R}[0pt]{0.51\textwidth}
    \begin{minipage}{\textwidth}
    \vspace{-2.5em}
    \begin{spacing}{0.8}
\begin{algorithm}[H]
    \SetAlgoNoLine
    \KwIn{s}
    $\itt{deque} \leftarrow [(s,\itt{null})]$ \\
    \While{$\itt{deque} \neq [] $}{
        $n, g \leftarrow \itt{deque}[0]$\\
        $\itt{deque} \leftarrow \itt{deque}[1..]$\\
        \If{$g = \itt{null}$}{
            do node computation on \itt{n}\\
            $g \leftarrow \itt{childGenerator}(n)$
        }
        $c \leftarrow \itt{g.yield()}$\\
        \If{$c \neq \itt{null}$}{
            $\itt{deque} \leftarrow [(c, \itt{null})] + \itt{deque} + [(n, g)]$
        }
    }
\caption{\algoShort~ Graph Search}
\label{algo:sot}
\end{algorithm}
\end{spacing}
    \end{minipage}
  \end{wrapfigure}

After this point Cain continues searching for shorter paths, and can cull any nodes with longer paths. During the search the same Goal-Bags may be reproduced in different ways, we cull the current node any time a Goal-Bag is produced that has already been seen at a lower or equal cost, or if the Goal-Bag has more Goals than available registers.

The second part of the search strategy defines the search order. Each invocation of the reverse search algorithm produces one new node, $\goalbag{C}$, and the input node is incremented to know how many of its children have been produced so far. Cain uses this simple definition to allow several graph traversal algorithms to be implemented. Using Depth-First-Search (DFS), Cain can simply maintain a stack of the nodes. On each cycle the top node is popped off the stack and given to the reverse search algorithm. Then the incremented parent node is put back on the stack, followed by the new child node.

While DFS performs well in AUKE, it struggles in Cain because the number of child nodes at every level is far greater, since each edge is only one instruction and there are multiple kernels to consider. This means the size of the graph we would like to search is much larger and we are unable to search even a small fraction of it. To overcome this we use a graph-traversal algorithm that, for our purposes, we call \algoName~ (\algoShort). The aim of this algorithm is to ensure that the search does not end up `trapped' in one small part of the graph, but can effectively search traverse many children of many of the nodes that are found where DFS will search all of the children of nodes at the extent of the paths it searches before searching the second children of nodes earlier in the graph. Algorithm \ref{algo:sot} shows a pseudo-code implementation of \algoShort. In each cycle the front of the queue is polled, if the node has not been seen before, Cain checks to see if it can be directly transformed from the initial-state Goal-Bag, this is the `node computation'.  The node is then passed to the reverse search algorithm to attempt to produce the next new child node and to increment parent node -- this is implicit in calling `\textit{yield()}' on \textit{g}. The child node, if it exists, is put on the front of the queue and the incremented parent node is put on the back.
We do not claim that \algoShort~ is novel, but we have found it superior to obvious alternatives, and the strategy used in~\cite{Linnea}; for details see \ifanonymous\else\cite{Stow2020}.\fi
\vspace{-1em}
\subsection{Cost Function}
In the reverse search algorithm we see that the pairs of $\goalbag{Uppers}$ and $\goalbag{Lowers}$ are produced one at a time. While this simplification allows us to produce more generic graph traversal implementations; what allows Cain to efficiently find solutions, are the heuristics that allow us to order the pairs that are produced for a node from the most promising to the least. This type of heuristic provides the order of siblings to search so we call it a `local heuristic'. It doesn't compare nodes in different parts of the graph, which we would call a `global heuristic'. We found that we were unable to find effective global heuristics because traversal algorithms that take advantage of such heuristics end up producing huge frontier sets of nodes making the memory requirements too large. The use of local heuristics drives the SCAMP-5 code generation in Cain instead, though support for best-first-search with global heuristics is available in Cain. The local heuristics used for SCAMP-5 are based on generating every child node of the parent and then ordering them based on a cost function. There are 3 main components considered for the cost: Atom distance, repeated Goals, and divisions. A simplified formula is shown in Equation \ref{eq:costFunc}.
\begin{align}
    cost(\goalbag{C}) &= \itt{dists}(\goalbag{C}) + \itt{reps}(\goalbag{C}) + \itt{divs}(\goalbag{C})\label{eq:costFunc}\\
    \itt{dists}(\goalbag{C}) &= \sum_{G\in \goalbag{C}}\left(|G|+ \sum_{a\in G} \left(  |a.x| + |a.y| \right) \times \begin{Bmatrix}
    \frac{1}{2} &\text{if} \not\exists B \in \goalbag{C}. G \subset B\\
    1 & \text{otherwise.}
    \end{Bmatrix}\right)\\
    \itt{reps}(\goalbag{C}) &= \sum_{\{G \in \goalbag{C}: \text{$G$ is unique wrt any translations} \}} \begin{Bmatrix}
    |G|^2 & \exists a,b \in G. a \neq b\\
    0 & \text{otherwise.}
    \end{Bmatrix}\\
    \itt{divs}(\goalbag{C}) &= \frac{2^d}{\text{min}(\itt{multiplicity}(a) \forall a \in G. \forall G \in \goalbag{C})}
\end{align}

The Atom distance part counts up how many Atoms every Goal in $\goalbag{C}$ has, and how far from the centre they are, with some relief if the Goal is a sub-Goal of another Goal in $\goalbag{C}$. The repeated Goals portion of the cost penalises $\goalbag{C}$ by the square of number of Atoms in each Goal, unless that Goal is equal to a translation of another Goal in $\goalbag{C}$. The divisions component penalises $\goalbag{C}$ for the number of division operations that would be required to produce the Goals from the identity-kernel Goal, $G_\itt{ID}$.

\section{Evaluation}\label{sec:evaluation}
All performance evaluation is conducted on an Intel Core i7-7700HQ CPU (4 cores, 8 threads) with a base frequency of 2.80GHz. The computer has 16GB of RAM, runs Ubuntu 18; as well as Java 1.8 (Oracle) and Python 3.6 to run Cain and AUKE respectively. The implementation of AUKE used, as developed by Debrunner, can be found on Github\footnote{\href{https://github.com/najiji/auto_code_cpa/tree/75c017e5ad28c0f3f040fb9f84d7f8727d035baa}{github.com/najiji/auto\_code\_cpa/tree/75c017e5ad28c0f3f040fb9f84d7f8727d035baa}}. 
\ifanonymous
\else
Cain source code can be found at \href{https://github.com/ed741/cain}{github.com/ed741/cain}, and the specific version and sources for experimental setups presented in this evaluation can be found at \cite{cain3.0-e.1-2020}.
\fi
\vspace{-1em}
\subsection{Performance Evaluation Against AUKE}
Comparison of our work Cain against AUKE is performed by comparing resulting code generated from the respective compilers, given the same input filters. Both compilers are given 60 seconds to find a solution using all 6 registers. Note as Cain supports multi-threading, it spawns 4 worker threads to perform the search. \ifanonymous \else  \fi

As shown in Table~\ref{table:kernelResults}, Cain significantly outperforms AUKE.
Cain supports a wider set of instructions in contrast of AUKE, enabling generation of more efficient code. Not only this, the search strategy used by Cain is better than AUKE's, as shown in $5\times5$ Gaussian Kernel, were using the same set of instructions (Basic), code generated by Cain is half in length when compared to output of AUKE's. Although, in further testing, AUKE is able to produce less inefficient code for this kernel given fewer registers. When given multiple kernels, Cain is able to perform simultaneous kernel optimisation. For example when combining $3\times3$ and $5\times5$ Gaussian, unlike AUKE, Cain is implemented to utilise the common sub-expressions between the kernels, thus, generating shorter code than naively concatenating the code for each of the Gaussian kernels. Neither Cain or AUKE perform a compete exhaustive search.

The AnalogNet2 filter is the kernels used in AnalogNet2~\cite{wong2020analognet}\cite{analognet2}, which is a CNN for SCAMP-5, capable of MNIST digit recognition. Cain requires only 21 instructions whereas AUKE produces kernel code which has in total 49 instructions. Reduced code not only improves the execution time, but also reduces the noise build up, which is significant problem as discussed in~\cite{wong2020analognet}. If the aim of finding sub-expressions is to eliminate redoing work, then the number of add and subtract operands is a proxy for how effective the search for sub-expressions is, regardless of how translations are handled. Table \ref{table:code} shows that AUKE's code has 40 add or subtract operands whereas Cain's code has only 27.
We have compared the runtime of AnalogNet2's convolution kernels, generated by AUKE and Cain on the physical SCAMP-5. Note, as AUKE produces code which performs invalid register manipulation, the fixed code as used in~\cite{analognet2}, which executes on the device is 81 instructions long.
The execution time of the code produced by AUKE and Cain for the convolution kernels were $35\mu\text{s}$ and $9\mu\text{s}$ respectively, showing almost 4 times speedup.

\begin{table}[h]
 \caption{Kernels tested in AUKE and Cain. Values on the righthand side of the table refer to the number of SCAMP-5 macro instructions in the programs generated by AUKE and  Cain for each filter. AUKE can only use the 'basic' macro instructions, so Cain is run twice; to compare its effectiveness under the same restrictions as AUKE. Since AUKE does not offer a way to compile multiple kernels at once, values for each kernel are given separately. \vspace{-0.5em}}
 \label{table:kernelResults}
 \centering
        \begin{tabular}{>{\centering\arraybackslash}m{6em} l >{\centering\arraybackslash}m{5em} >{\centering\arraybackslash}m{2.5em} >{\centering\arraybackslash}m{2.5em}}
            \toprule
            Name & Approximated Filter & AUKE & \multicolumn{2}{c}{Cain} \vspace{-0.2em}\\
            \cmidrule{3-5}
            \vspace{-0.2em}
                 &                     & Basic & All & Basic \\
\midrule
3$\times$3 Gauss & $\left\{
\frac{1}{16}\begin{bsmallmatrix} 1 & 2 & 1 \\ 2 & 4 & 2 \\ 1 & 2 & 1 \\ \end{bsmallmatrix}
\right\}$ \vspace{0.1em}&12&\textbf{10}&12 \\ 
5$\times$5 Gauss & $\left\{
\frac{1}{64}\begin{bsmallmatrix} 0 & 1 & 2 & 1 & 0 \\ 1 & 4 & 6 & 4 & 1 \\ 2 & 6 & 10 & 6 & 2 \\ 1 & 4 & 6 & 4 & 1 \\ 0 & 1 & 2 & 1 & 0 \\ \end{bsmallmatrix}
\right\}$\vspace{0.1em} & 50&\textbf{19}&25\\ 
5$\times$5 and 3$\times$3 Gauss & $\left\{
\frac{1}{64} \begin{bsmallmatrix} 0 & 1 & 2 & 1 & 0 \\ 1 & 4 & 6 & 4 & 1 \\ 2 & 6 & 10 & 6 & 2 \\ 1 & 4 & 6 & 4 & 1 \\ 0 & 1 & 2 & 1 & 0 \\ \end{bsmallmatrix},
\frac{1}{64} \begin{bsmallmatrix} 0 & 0 & 0 & 0 & 0 \\ 0 & 4 & 8 & 4 & 0 \\ 0 & 8 & 16 & 8 & 0 \\ 0 & 4 & 8 & 4 & 0 \\ 0 & 0 & 0 & 0 & 0 \\ \end{bsmallmatrix}
\right\}$\vspace{0.1em} &$(50 + 12)$&\textbf{26}&39\\ 
AnalogNet2 & $\left\{\begin{array}{c}
    \frac{1}{4} \begin{bsmallmatrix}  0 &  0 & 0 \\ -3 & 1 &  0 \\ -3 &  0 & 2 \\ \end{bsmallmatrix},
    \frac{1}{4} \begin{bsmallmatrix} -4 & -1 & 1 \\ -1 & 2 &  0 \\  1 &  1 & 0 \\ \end{bsmallmatrix}, \vspace{0.3em}
    \frac{1}{4} \begin{bsmallmatrix} -1 &  2 & 0 \\ -1 & 1 & -3 \\  0 & -3 & 0 \\ \end{bsmallmatrix}
    \end{array}\right\}$ &$\begin{array}{c}
         (13 + 21\\ + 15) 
    \end{array}$&\textbf{21}&30\\ 

            \bottomrule
        \end{tabular}
        \vspace{-0.5em}

\end{table}

\begin{table}[thb]   
    \caption{Comparison of Code for the AnalogNet2 filter generated by AUKE and Cain. The Input Register is `A' and the output registers for the 3 kernels are `A',`B',`C' respectively. For AUKE, kernel 2 is run first since testing showed it was longest so this gives AUKE more registers to use.\vspace{-1em}}
    \label{table:code}
    \centering
    \begin{tabular}{p{2.7cm}p{2.7cm}p{2.7cm}p{3.5cm}}
        \toprule
        \multicolumn{3}{c}{AUKE} & \multicolumn{1}{c}{Cain} \\
        \hline
\vspace{-1em}
\tiny Kernel 2
\vspace{-1em}
\begin{lstlisting}[numbers=left,xleftmargin=2.5em,basicstyle=\ttfamily\tiny]
mov(B,A);
divq(B,B);
divq(B,B);
movx(C,B,north);
neg(C,C);
neg(D,C);
movx(E,D,west);
neg(E,E);
add(F,B,E);
movx(B,D,east);
add(B,B,E);
movx(D,E,south);
movx(D,D,south);
sub(B,B,D);
add(B,B,F);
add(B,C,B);
movx(C,C,west);
add(B,B,C);
movx(C,F,south);
add(B,C,B);
add(B,B,F);
\end{lstlisting} &
\vspace{-1em}
\tiny Kernel 3
\vspace{-1em}
\begin{lstlisting}[numbers=left,firstnumber=22,xleftmargin=2.5em, basicstyle=\ttfamily\tiny]
mov(C,A);
divq(C,C);
divq(C,C);
movx(D,C,south);
neg(D,D);
movx(E,C,east);
sub(D,D,E);
movx(E,C,north);
add(E,E,D);
add(D,D,D);
add(D,E,D);
movx(E,C,west);
sub(C,C,E);
add(D,D,C);
movx(C,C,north);
add(C,D,C);
\end{lstlisting} &
\vspace{-1em}
\tiny Kernel 1
\vspace{-1em}
\begin{lstlisting}[numbers=left,firstnumber=38,xleftmargin=2.5em, basicstyle=\ttfamily\tiny] 
divq(A,A);
divq(A,A);
movx(D,A,west);
neg(D,D);
movx(E,D,south);
add(D,D,E);
add(E,A,D);
movx(A,A,south);
movx(A,A,east);
add(A,D,A);
add(A,A,A);
add(A,E,A);
\end{lstlisting} &
\vspace{-1em}
\begin{lstlisting}[numbers=left,xleftmargin=2.5em,basicstyle=\ttfamily\tiny]
diva(A,D,E);
div(D,E,C,A);
movx(E,D,west);
movx(C,E,north);
neg(F,E);
subx(B,F,east,A);
addx(E,E,D,south);
add2x(D,F,D,north,north);
sub2x(F,D,south,south,C);
add2x(D,C,D,east,south);
add(E,E,D);
movx(D,A,north);
add2x(A,C,A,east,east);
movx(C,B,east);
add(D,F,D);
add2x(F,F,E,east,south);
movx(E,B,south);
addx(A,B,A,south);
addx(A,B,A,west);
add2x(B,F,B,north,west);
add(C,D,C,E);
\end{lstlisting}
    \end{tabular}
    \vspace{-2em}
\end{table}

\subsection{Effectiveness of the Search Strategy}
If Cain has an effective heuristic we will quickly see a point of diminishing returns in code length, as Cain continues to search new nodes and takes more time.
We can track the number of nodes that are explored before finding any plan in Cain, and so use this as a measure of the search strategy and heuristics that is more independent of physical compute performance. With this in mind we test the effectiveness of our heuristic by constructing 100 samples of randomly generated single kernel filters as in Equation~\ref{eq:randomfilter1}. Running Cain as per the following configuration -- Maximum Nodes to Explore: 20000, Maximum Search Time: 60s, Worker Threads: 1 -- allows us to collect as many plans as can be found in the given time limit. We then ran Cain again, but with Cain's SCAMP-5 heuristic disabled and replaced with a random sort. This allows us to compare Cains heuristics against an unaided benchmark.
\begin{equation}
\label{eq:randomfilter1}
\begin{array}{c}
    \frac{1}{8}
    \begin{bsmallmatrix}
    u_1 & u_2 & u_3 \\
    u_4 & u_5 & u_6 \\
    u_7 & u_8 & u_9 
    \end{bsmallmatrix} 
    \\
    \text{ \footnotesize Given $u_1 .. u_9$ are integers sampled uniformly from the range $[0..8]$}
    \end{array}
\end{equation}

\begin{figure}[tb]
    \centering
    \includegraphics[width=0.49\textwidth ]{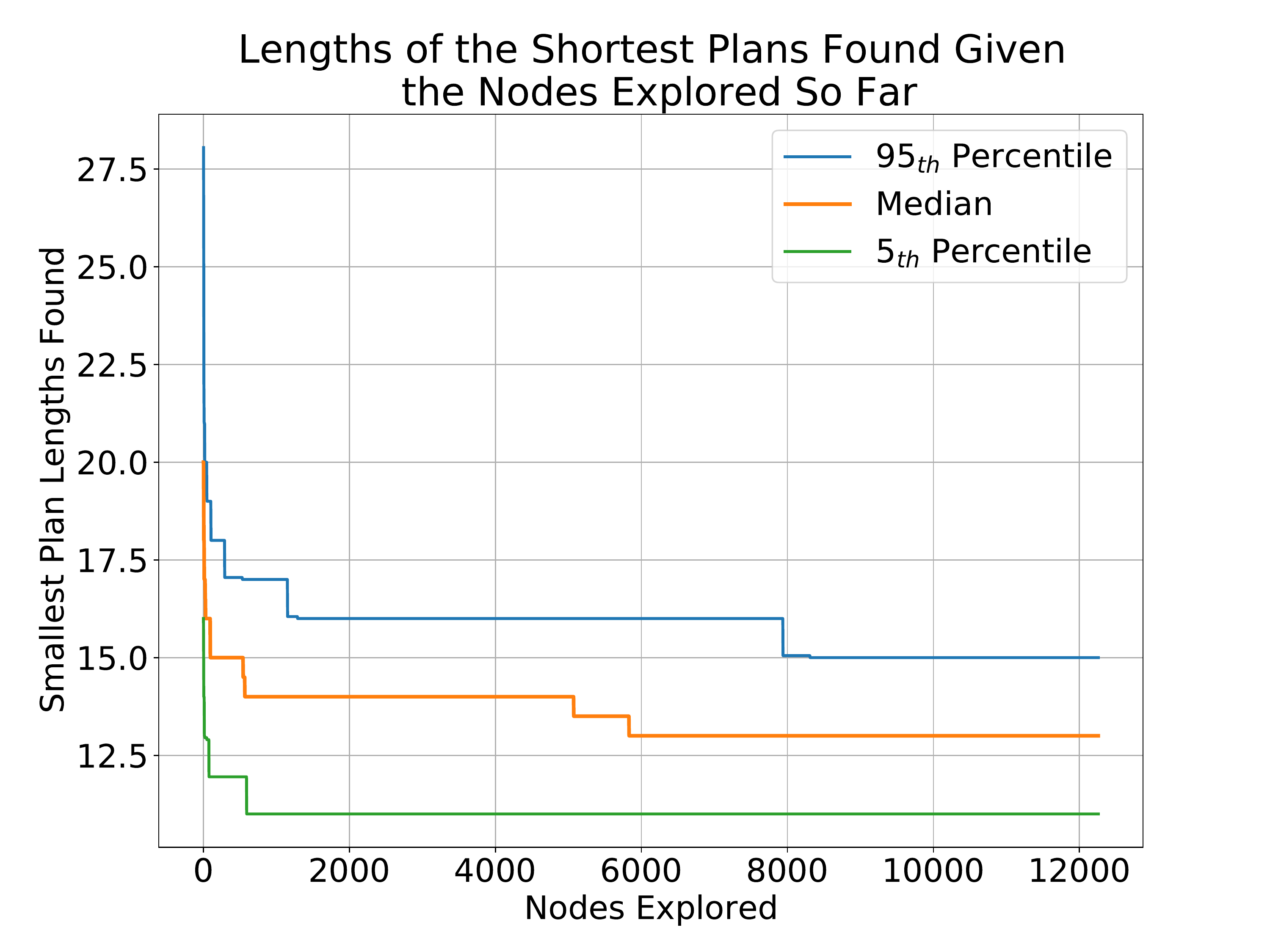}
    \includegraphics[width=0.49\textwidth ]{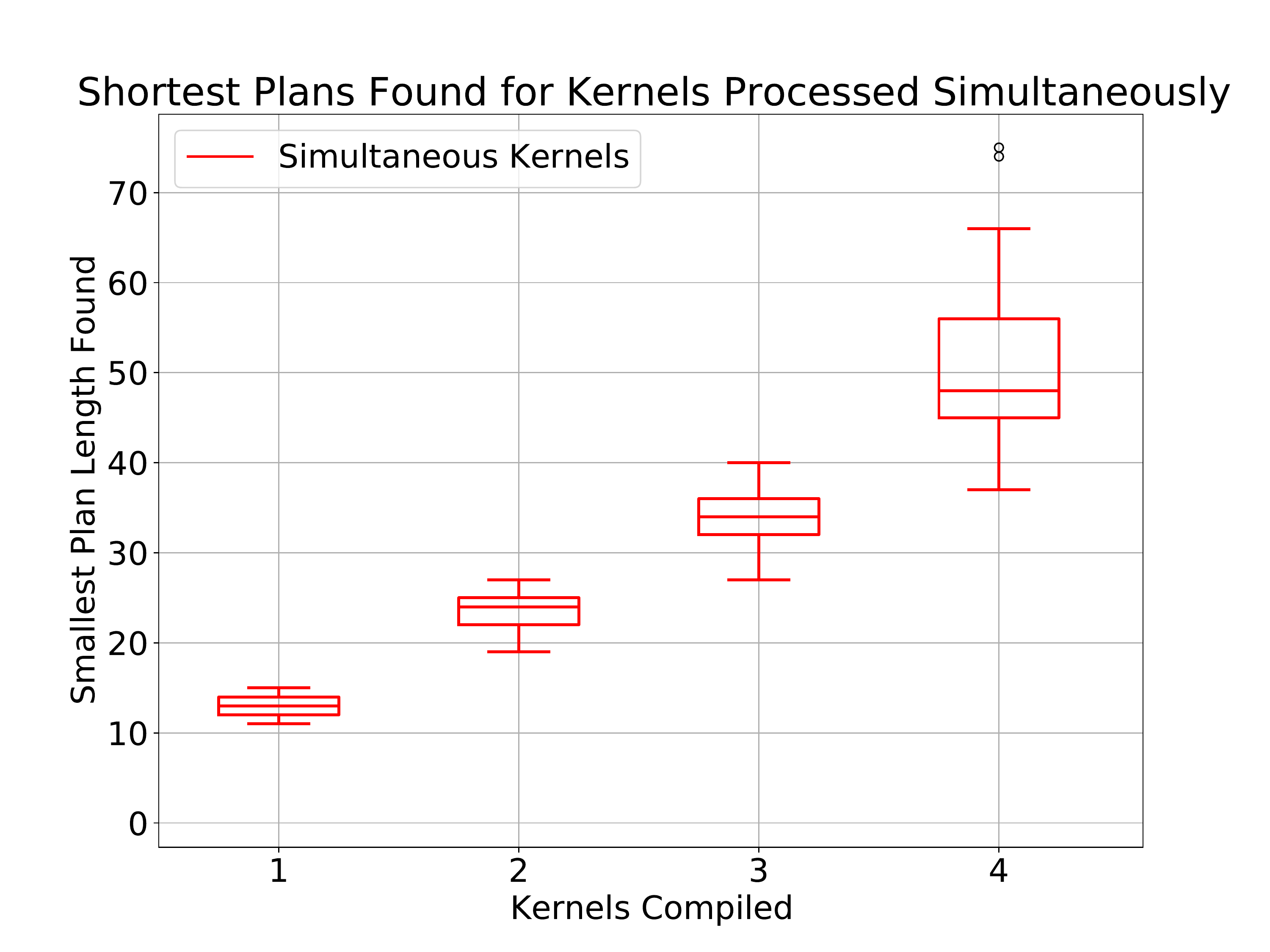}
    \caption{Left: Graph showing the median number of instructions in the best plans found before $n$ nodes have been explored by Cain. With 100 samples of randomly generated singular $3 \times 3$ kernel filters. Right: Graph showing the number of instructions in the shortest programs found by Cain for filters with 1, 2, 3, and 4 random $3\times 3$ kernels. 25 samples were produced for each kernel count.\vspace{-0.5em}}
    \label{fig:NodesExplored_MultiKernel}    
    \vspace{-0.5em}
\end{figure}

 We found that Cain was unable to find any plan for any of the 100 sample filters without its heuristics, principally demonstrating that effective heuristics are required in Cain for any tangible progress to be made. We plot the lengths of the best plans found against the number of nodes expanded before the plan is found in Fig. \ref{fig:NodesExplored_MultiKernel}. We can see that improvements are fewer and further between after the first 2500 nodes are explored. After this we see that we can expect at most a reduction equal to the reduction seen at 2500 for the rest of the nodes explored. This clearly demonstrates a point of diminishing returns for these filters. If the heuristic is effective we expect it to direct the search towards short plans first, and try instructions less likely to be optimal later. This model fits the data well as we see short plans are found quickly, and while improvements can be made, it is clear that they are found less often as the search continues.

\subsection{Effectiveness of the Simultaneous Kernel Optimisation}
One of the significant features of Cain is to efficiently generate code for filters with multiple kernels, and do this simultaneously such that shared common sub-expressions can be reused. As it is possible for Cain to perform exhaustive searches for plans, given sufficient time, it will find a solution that simply computes the individual kernels independently, or find a solution with lower cost -- utilising the common sub-expressions.

First, we wish to test whether the length of generated code is sub-linear to the number of input kernels.
To test this, we again generate kernels using the using the method in Equation \ref{eq:randomfilter1}. For kernel counts from 1 to 4 we generated 25 filters each and test them all using the same configuration as before except that we remove the maximum nodes explored constraint, and allow 4 worker threads. We plot the results in Fig. \ref{fig:NodesExplored_MultiKernel} and see that the results appear worse than linear, suggesting that common sub-expressions are not effectively being taken advantage of.

We hypothesise that the limited number of registers in the SCAMP-5 architecture is the major limiting factor in producing efficient code. To test this we increase the number of available registers to 18. For filters with 1 kernel up to 10 kernels we generate 10 samples each. Every kernel in the 100 filters is produced as in Equation \ref{eq:randomfilter1}. For each sample, Cain compiles the kernels individually, given the appropriate number of registers such that other kernels in the filter would not be overwritten. Then we compile the kernels simultaneously using Cain. All compilations are given 60s to run, with 4 worker threads. 

\begin{figure}[tb]
    \centering
    \includegraphics[width=\textwidth, trim={0 0 0 3.3cm},clip]{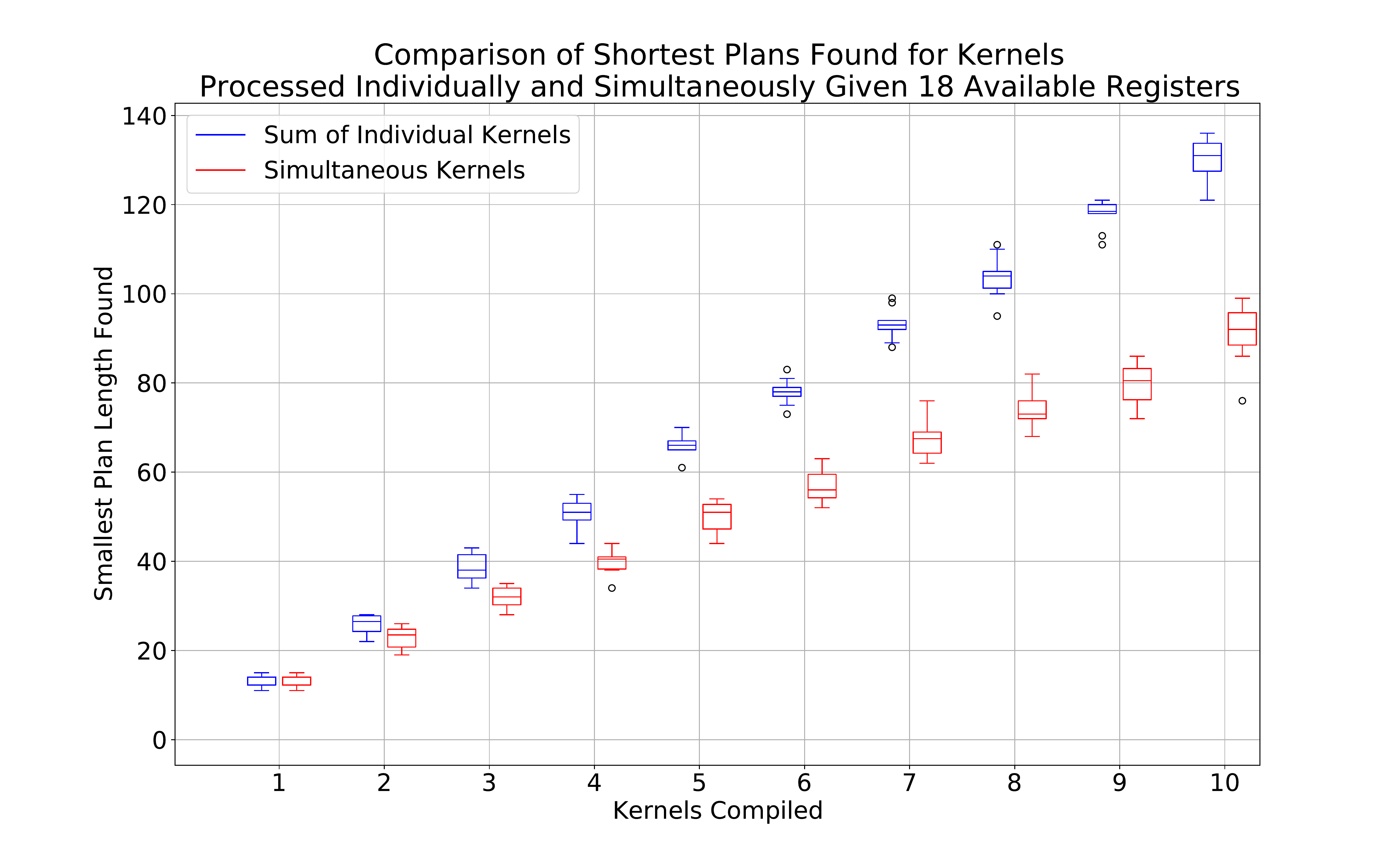}
    \caption{Graph comparing the sum of the shortest SCAMP-5 code lengths found for kernels compiled individually, against the same kernels compiled simultaneously as one filter. For each filter a total of 18 registers were made available (more than in SCAMP-5) to reduce register availability as a limiting factor. In total 100 filters are produced, 10 for each number of kernels per filter. Each kernel is a randomly generated $3 \times 3$ kernel with coefficients uniformly selected in eighths from 0 to 1 (inclusive).\vspace{-0.5em}}
    \label{fig:SmallestPlanLengthOverKernelCountComparason24Reg}
    \vspace{-1em}
\end{figure}

Fig. \ref{fig:SmallestPlanLengthOverKernelCountComparason24Reg} shows the results of this test. We see clearly that when register limitations are not a restricting factor Cain is able to consistently improve the performance of filter implementations by compiling them simultaneously.
We see that improvements grow with more kernels, and it appears that the length of code generated for simultaneously compiled kernels increases sub-linearly. This supports the idea that with more kernels, ever more common-sub expressions can be exploited.

\section{Related Work: AUKE}\label{sec:related_work}

In this section we look at how AUKE operates to provide extra context and contrast for Cain. Automatic Kernel Code Generation for Analogue SIMD (AUKE) is an algorithm for generating code given a single convolutional kernel created by T. Debrunner~\cite{TomD}. It can be characterised by 4 main steps: kernel approximation; the reverse split algorithm; graph relaxation; and finally register allocation. First, AUKE approximates the input kernel into the Goal representation. In this process Cain is similar to AUKE and the reasoning and mechanics have been discussed in Section \ref{sec:cain:goalDef}.

Unlike in Cain, multiple instructions are represented by a single elemental transformation of Goals. These elemental transformations form edges of a graph that describe the translation, addition, subtraction and division of Goals to produce the desired convolutions filter. This abstraction allows AUKE to reduce the effective size of the search space at the cost of granularity in instruction selection and being extensible to hardware features such as 3-operand addition. Debrunner called this the `Reverse-Split Algorithm'.

The graph of elemental transformations is dynamically generated via a recursive depth-first search that tries to split a Goal $G$, that needs to be produced, into 3 sub-Goals:
\vspace{-0.5em}
\begin{equation}
    G = U \cup L \cup R \label{eq:AukeReverseSplit}\quad \text{ where } U = \itt{elementalTransformation}(\itt{L})
\end{equation}\vspace{-1.7em}

This recursive algorithm then means that if the search can find solutions for $L$ and $R$ (two smaller problems) it can trivially create $U$ and therefore the desired Goal.

In the ideal case $R=\emptyset$ and so only $L$ needs to be produced and we save one addition. In the worst case $L = U = \emptyset$ and $R$ is a transformation of $G$ and so less useful work is done in that step. If two Goals are equal they are merged such that they aren't calculated twice, to exploit common sub-expressions in the Goals. This process is repeated until a single Goal, the initial-Goal, is left. This algorithm is able to entirely search the relevant problem space, given a couple of assumptions. Most notably, the assumption that every sub-Goal generated is a subset of the Final-Goal. This reduces the search space significantly to the most promising but not necessarily the best solutions, allowing AUKE to find generally effective solutions.

The algorithm is made efficient and useful by intelligently selecting the order with which $U$s, $L$s, and $R$s are generated at every recursive step. By selecting pairs of $U$ and $L$ that are likely to lead to efficient code, the algorithm can quickly find some path to the initial-Goal. From then on the recursive search can stop early if a lower cost solution has already be found.

The Graph Relaxation step aims to mitigate missing optimal solutions because of the assumption that sub-Goals are always subsets of the Final-Goal by using a `retiming' algorithm used in integrated circuit design. This is not needed in Cain since Cain searches instruction by instruction, and so any optimisations found via graph relaxation are already a part of the search space. 

The final step is to perform register allocation on the graph to be able to generate usable code. A maximum bound of registers is already accounted for in the search algorithm, since spilling is not an option for the SCAMP-5 architecture. For this task; variable liveness is considered for each node of the graph representation, and a graph colouring algorithm is used to find a solution.

\section{Conclusion}
\label{sec:conclusion}
We have presented Cain, a compiler which produces SCAMP-5 instructions from a set of convolutional kernels.
Although the effectiveness of simultaneous kernel optimisation is limited on the current iteration of the SCAMP-5, we demonstrate, that with the increased number of registers, the length of the output of Cain is sub-linear to the number of kernels given.
We have conducted extensive comparison against AUKE, and we demonstrate that the code generated by Cain is more efficient, and exhibits almost 4x speed up when the generated kernel is executed on the SCAMP-5 device.
We believe that SCAMP-5 is a strong candidate for edge computation, and by providing easy to use, yet efficient code generation toolkit, we hope to accelerate the relevant research in this field.

\paragraph*{Acknowledgements}
\ifanonymous
\else
We would like to thank Piotr Dudek, Stephen J. Carey, and Jianing Chen at the University of Manchester for kindly providing access to SCAMP-5, and their support in our work.  This work was partially supported by the EPSRC, grant reference EP/P010040/1.
\fi

\bibliographystyle{splncs04}
\bibliography{bib.bib}

\end{document}